\documentstyle[12pt, epsf]{article}
\def\gtwid{\mathrel{\raise.3ex\hbox{$>$\kern-.75em\lower1ex\hbox{$\sim$}}}}
\def\ltwid{\mathrel{\raise.3ex\hbox{$<$\kern-.75em\lower1ex\hbox{$\sim$}}}}
\def\gev{{\rm \, Ge\kern-0.125em V}}
\def\tev{{\rm \, Te\kern-0.125em V}}
\def\sn{\widetilde\nu}
\def\msn{m_{\sn}}
\def\sel{{\widetilde e}_{\scriptscriptstyle\rm L}}
\def\seli{{\widetilde e}_{\scriptscriptstyle{\rm L}i}}
\def\selj{{\widetilde e}_{\scriptscriptstyle{\rm L}j}}
\def\msel{m_{\sel}}
\def\Hone{H^0_1}
\def\Htwo{H^0_2}
\def\Hthree{H^0_3}
\def\mH3{m_{\Hthree}}
\def\beq{\begin{equation}}
\def\eeq{\end{equation}}

\begin{document}
\begin{titlepage}
\pagestyle{empty}
\baselineskip=21pt
\rightline{hep-ph/9409270}
\rightline{UMN--TH--1306/94}
\rightline{UCSBTH--94--29}
\rightline{September 1994}
\vskip1.25in
\begin{center}
{\large{\bf
Heavy Sneutrinos as Dark Matter}}
\end{center}
\begin{center}
\vskip 0.5in
{Toby Falk,$^1$ Keith A.~Olive,$^2$  and Mark Srednicki$^1$
}\\
\vskip 0.25in
{\it
$^1${Department of Physics,
University of California, Santa Barbara, CA 93106, USA}\\
$^2${School of Physics and Astronomy,
University of Minnesota, Minneapolis, MN 55455, USA}
}\\
\vskip 0.5in
{\bf Abstract}
\end{center}
\baselineskip=18pt \noindent
We calculate the relic density of very heavy, stable scalar neutrinos in the
minimal supersymmetric standard model.  We include all two-body final states,
as well as the effects of co-annihilation with scalar electrons.  We find that
the sneutrino relic density is in the cosmologically interesting region
$0.1\ltwid\Omega_{\sn}h^2\ltwid 1.0$ for $550\gev\ltwid\msn\ltwid 2300\gev$.
For nominal values of the parameters of our galactic halo, recent results from
the Heidelberg--Moscow direct detection experiment set an upper limit on
$\Omega_{\sn}$ which is lower by a factor ranging from two to ten,
depending on $\msn$.
\end{titlepage}
\baselineskip=18pt

A sneutrino which is the lightest supersymmetric particle (LSP) is a
theoretically attractive dark matter candidate \cite{hkr,iban}.
Previous studies of sneutrino relic densities \cite{hkr,iban} have been
concerned with a sneutrino with a mass $\msn$ that is less than $m_W$.
Heavier sneutrinos can annihilate into a much greater set of final states;
for $\msn \simeq 100\gev$, the annihilation rate is very large, and
the relic density of sneutrinos is much too small to be of cosmological
or astrophysical interest.  However, as the sneutrino mass is increased,
the annihilation rate drops and the relic density rises.  We show that
heavy ($\msn\gtwid 500\gev$) sneutrinos may have a cosmologically interesting
relic density.  For nominal values of the parameters of our galactic
halo, direct detection experiments \cite{ahlen} - \cite{heidmos}
already place constraints
on this range of sneutrino masses, and
recent results from the Heidelberg--Moscow direct detection
experiment \cite{heidmos} may even rule out the values of
$\Omega_{\sn}\equiv\rho_{\sn}/\rho_{\rm crit}$ that we find.

The calculation of
relic densities of heavy sneutrinos involves an additional complication
not present for light sneutrinos, arising from the need to
consider coannihilations \cite{gs} with selectrons.
The masses of the sneutrino $\sn$ and left-handed selectron $\sel$
are related via \cite{hk}
\begin{equation}
\msel^2 = \msn^2 - m_Z^2 \cos{2\beta} \cos^2{\theta_{\rm W}} \;,
\end{equation}
where $\tan \beta$ is the ratio of the vacuum expectation
values of the two Higgs doublets.  We see that for
$\tan\beta > 1$ we have $\msel > \msn$, as we require if the sneutrino
is to be the LSP.  However, for $\msn\gg m_W$, the mass splitting
between $\sn$ and $\sel$ will be small.
Consequently their number densities are approximately equal until well
after the sneutrinos go out of chemical equilbrium with all other particles,
and we must follow the selectron density as accurately as the sneutrino
density.

We compute the tree level cross-section for annihilation to all allowed
two particle final states.
Table 1 lists the set of final states we consider.  The $f_d$ and $f_u$
refer to $T_3 = -1/2$ and $T_3 = +1/2$ fermions, respectively.
As intermediate states
we allow sneutrinos, selectrons, the gauge bosons $\gamma, W^\pm$ and $Z^0$,
the three Higgs bosons of the minimal
supersymmetric standard model (MSSM), $H_i^0$ ($i=1,2,3$), the charginos
$\widetilde\chi_i^\pm$ ($i=1,2$), and the neutralinos $\widetilde\chi_i^0$
($i=1,2,3,4$).  There are also four-scalar point interactions.
The couplings are those of the MSSM \cite{hk,gh}.
For simplicity, we ignore sfermion (in particular selectron) mixing, and
for convenience we
use the relation $M_1 = {5\over3} M_2 \tan^2\theta_{\rm w}$ between the
$U(1)$ gaugino mass $M_1$ and the $SU(2)$ gaugino mass $M_2$ which follows
from grand unification.  At tree level, the
masses of the scalar Higgs particles $\Hone$ and $\Htwo$ can be
written in terms of $m_Z, \tan\beta, $ and the mass the pseudoscalar
$\Hthree$.  (Although the one-loop corrections to the Higgs masses
can be large, they will not impact our results significantly, and
so we do not include them.)
We are then left with five free parameters: $\tan\beta$,
$\mH3$, $\msn$, $M_2$, and the Higgs mixing parameter $\mu$.
Since our results are largely insensitive to $\tan\beta$, we fix
$\tan\beta=2$ in our numerical work.

\begin{table}[htb]\caption{Initial and final states}
\begin{center}
\begin{tabular}{c|l}\hline
Initial State & Final States\\ \hline
\\
$\sn\sn^*$    & $f\bar f\, ,\, ZZ\, ,\, W^+W^-, $\\
              & $Z H_{(1,2,3)}^0\, ,\, W^\pm H^\mp\, ,\, H^+H^-,$ \\
              & $H_{i}^0 H_{i}^0 (i=1, 2, 3)\, ,\, H_1^0 H_2^0\, ,\,
H_{(1,2)}^0 H_3^0$\\
\\
$\sel\sel^*$  & same as above, plus $\gamma\gamma\, ,\,\gamma Z$\\
\\
$\sn\sel^*$   & $Z W^+\, ,\,\gamma W^+\, ,\, Z H^+\, ,\,\gamma H^+\, ,\,
                W^+ H_{(1,2,3)}^0 ,$\\
              & $H^+ H_{(1,2,3)}^0\, , \, f_u \bar f_d$\\
$\sn\sn$      & $\nu\nu$\\
$\sel\sel$    & $e e$\\
$\sn\sel$     & $\nu e$\\

\end{tabular}
\end{center}
\end{table}

To derive a thermally averaged
cross section, we make use of the technique of ref.~\cite{swo}.  We expand
$\langle\sigma v_{\rm rel}\rangle$ in a Taylor expansion in powers of
$x = T/\msn$:
\beq
\langle \sigma v_{\rm rel } \rangle = a + b x + O(x^2)\;.
\eeq
Repeating the analysis of ref.~\cite{swo} for initial particles with
different masses $m_1$ and $m_2$ yields
\beq
\langle\sigma v_{\rm rel}\rangle = \frac{1}{m_1m_2}\left(1 -
 \frac{3(m_1 + m_2)T}{2 m_1m_2}\right) w(s)|_{s\to (m_1+m_2)^2 + 3(m_1+m_2)T}
\; +\; O(T^2).
\label{e}
\eeq
\noindent Here
\begin{eqnarray}
w(s) & = & \frac{1}{4}\int\; d{\rm LIPS}\; |{\cal M}|^2\\
\noalign{\medskip}
     & = & \frac{1}{32\pi}\frac{p(s)}{s^{1/2}}\int^{+1}_{-1}
            d\cos\theta_{\rm CM}\;|{\cal M}|^2,
\end{eqnarray}
\noindent where $d{\rm LIPS}$ is the Lorentz Invariant Phase Space element,
$p(s)$ is the magnitude of the three-momentum of one of the initial particles
in the center-of-mass frame as a
function of the total center-of-mass energy squared $s$,
$\theta_{\rm CM}$ is the center-of-mass scattering angle, and
$|{\cal M}|^2$ is the absolute square of the reduced matrix element for
the annihilation, summed over final spins.  The $a$ and $b$ coefficients
may be read off the right-hand side of Eq.$\;$\ref{e} after expanding
in powers of $x$.

The sneutrinos and selectrons remain in chemical equilibrium with each
other through freeze-out.  We consider the total
density $n = \sum_i n_i$, where the index runs over $\sn, \sn^*, \sel,$
and $\sel^*$.  We write the rate equation for $n$:
\beq
\frac{dn}{dt}  =  -3Hn - \langle \sigma_{\rm eff} v_{\rm rel } \rangle
                    (n^2 - n_{\rm eq}^2).
\label{b}
\eeq
\noindent Here
\beq
\sigma_{\rm eff} = \sum_{ij}\sigma_{ij} r_i r_j,
\eeq
\noindent where $H$ is the  Hubble parameter,
$r_i = n_i^{\rm eq}/n^{\rm eq}$, $n^{\rm eq}_i$ is the equilibrium
density of particle species $i$,
and $\sigma_{ij}$ is the
total cross-section for the $i^{\rm th}$ particle to annihilate with the
$j^{\rm th}$ particle.  Long after freeze-out, the selectrons decay into
sneutrinos, leaving a number density $n$ of sneutrinos.
We rewrite the derivative in eq.~(\ref{b}) in terms of the temperature $T$
and numerically integrate the resulting equation down to $T=2.75\; {\rm K}$
to get the relic number (and hence mass) density of sneutrinos.

We find that the dominant contribution to $\sigma_{\rm eff}$ comes from
annihilations of $\sn{\sn}^*$, $\sel\sel^*$, and $\sn\sel^*$ to gauge bosons,
and $\sn\sn$, $\sel\sel$, and $\sn\sel$ to neutrinos and electrons via
neutralino and chargino exchange.

In fig.~1 we show the relic density $\Omega_{\sn}h^2$ as a function of
sneutrino mass; $h$ is the Hubble parameter today in units of
$100\,$km/s/Mpc.  The dashed line assumes that two of the sneutrino
generations are much heavier than the third, and that the lightest
neutralino is just 10\% heavier than the sneutrino.  (We set $\mu=M_1$,
but since the higgsinos do not couple significantly to the sneutrinos,
the value of $\mu$ does not affect our results.  Also, the
pseudoscalar Higgs mass was taken to be $\mH3 = 500\gev$, but
changing this parameter also has a negligible effect.)
Note that  $\Omega_{\sn}h^2$ passes
through the cosmologically interesting region $0.1\ltwid \Omega_{\sn}
h^2 \ltwid 1.0$ for $550\gev\ltwid \msn \ltwid 2300 \gev$.
Increasing $M_1$ raises the masses of the gauginos
and eventually turns off the gaugino-mediated annihilations of
$\sn\sn$, $\sel\sel$, and $\sn\sel$ to neutrinos and electrons.
The dot-dashed line in fig.~1 is for this case, with
$M_1 \simeq 10\,\msn$.
We can also consider the case where all three generations of sneutrinos
have very similar masses, and so are all in chemical equilibrium with
each other at the time of freeze-out.  Although one might expect that
this case simply results in a value of $\Omega_{\sn} h^2$ which is three
times larger, in fact it is a little more complicated because some
annihilation channels ($\sn_i\sn_j\to\nu_i\nu_j$,
$\sn_i\selj\to\nu_i e_j$, $\seli\selj\to e_i e_j$) can now occur in
nine different ways, whereas all the others ($\sn_i\sn_i^*\to f\bar f$, etc.)
can occur in three different ways.  In fig.~1, the solid line corresponds
to three equal-mass generations of LSP sneutrinos, but with
the neutralinos again as light as possible ($M_1 = \mu \simeq 1.10\,\msn$).
When the gauginos are heavy, then $\Omega_{\sn}h^2$ is very close to
what it would be for one generation of LSP sneutrinos.

The choice of masses for the sfermions here may seem a bit ad hoc.  In
the simplest grand unified
model where all the sfermions share a common scalar mass
parameter $m$, radiative corrections drive the mass of the sneutrino
above the mass of the right-handed selectron \cite{bor}.  To keep the
sneutrino as the lightest sfermion, and to prevent the sneutrino mass
from being driven above the mass the lightest neutralino,
we must invoke new high-energy physics to alter
the running of the sneutrino mass.  The GUT relation between $M_1$ and
$M_2$ may then be affected, but we have seen that the effect of
changing the gaugino masses by a factor of a few is small.

Non-detection of halo particles in direct detection experiments have
put bounds on the mass and relic abundance of dark matter sneutrinos
\cite{ahlen,druk,caldwell,collar,reuss,heidmos}.  The dotted line in fig.~1
marks the most stringent reported upper bound \cite{heidmos} on the relic
density $\Omega_{\sn}$ of sneutrinos for a given sneutrino mass $\msn\,$.
We assume the total dark matter density $\Omega_{\rm DM} = 1$, so that
the fraction of the dark matter halo which is composed of sneutrinos
is $\Omega_{\sn}$.  As in \cite{heidmos}, we take the local halo density of
dark matter to be $0.3 \gev/{\rm cm}^3$;  then
$\rho_{\sn} = (0.3 \gev/{\rm cm}^3)\Omega_{\sn}$.
The cross-section for sneutrino-nucleus interactions
is four times greater than the cross-section for Dirac neutrino-nucleus
interactions \cite{gw,sos} (not two times greater, as sometimes stated
\cite{ahlen,collar}).  In fig.~1, we take the published data of
\cite{reuss} to determine the upper bound on $\Omega_{\sn}$. We compare their
cross section constraint to four times the Dirac neutrino cross section.
We see that there are upper
bounds of $\msn\ltwid 600 \gev$ and $\Omega_{\sn}h^2 \ltwid 0.08\gev$
(for $h=1$).  Most cosmologically interesting relic densities are excluded.
This assumes that the dark matter which is not composed
of sneutrinos interacts less strongly with direct detection experiments
than do sneutrinos.  Also in fig.~1, the lower dotted line is found from
$\Omega = \sigma_{\rm b}/4\sigma_{\rm c}$, where
$\sigma_{\rm b}$ and $\sigma_{\rm c}$ refer to the curves from fig.~3
of \cite{heidmos}.  From this experiment we see that our computed range
of $\Omega_{\sn}$ is ruled out by
a factor ranging from roughly two to ten (for $h=1$).
If we consider sneutrinos as the sole component of the dark matter,
then we should fix $\rho_{\sn} = 0.3 \gev/{\rm cm}^3$ independent of
$\Omega_{\sn} h^2$; in this case we get a {\it lower} bound on the
sneutrino mass of $\msn \gtwid 17 \tev$, which we find requires a value of
$\Omega_{\sn}h^2$ that is much bigger than one.
However, we should keep in mind that the uncertainty in the local
value of our halo density is at least a factor of two \cite{flores}.
Furthermore, the momentum transfer in the interaction of a heavy sneutrino
with a detector nucleus is large; a calculation of the interaction
cross-section
must therefore include nuclear form factors, which are not well known.

To summarize, we have calculated the relic density of sneutrinos in a
model where a heavy ($\gtwid 500 \gev$) sneutrino is the LSP.  We have
found that sneutrino masses in the range $550\gev \ltwid \msn \ltwid
2300\gev$ may give cosmologically interesting relic densities, but that
this range is ruled out for nominal values of the parameters of our
galactic halo by a factor ranging from roughly two to ten.  Uncertainties
in the halo parameters and nuclear form factors still leave a marginal
chance for the heavy sneutrino to be a component of the halo dark matter.

\bigskip
\vbox{
\noindent{ {\bf Acknowledgements} } \\
\noindent  This work was supported in part by DOE grant DE--FG02--94ER--40823
and NSF grant PHY--91--16964.}


\newpage
\noindent{\bf{Figure Caption}}

\vskip.3truein

\begin{itemize}
 \item[]
\begin{enumerate}
\item[]
\begin{enumerate}

\item[Fig.~1)]The relic density $\Omega_{\sn}h^2$ as a function of $\msn$.
The dashed line corresponds to $\mH3 = 500\gev$ and
$m_{\widetilde\chi^0_1} \simeq 1.10\,\msn$, with one generation of LSP
sneutrinos.  The dot-dashed line is for
$m_{\widetilde\chi^0_1} \simeq 10\,\msn$.
The solid line corresponds to the same parameters as the dashed line,
but with three generations of LSP sneutrinos.  The Neuchatel-CIT \cite{reuss}
and the Heidelberg--Moscow \cite{heidmos} direct
detection experiments exclude values of $\Omega_{\sn}$ (not $\Omega_{\sn}h^2$)
above the upper and lower dotted lines respectively.

\end{enumerate}
\end{enumerate}
\end{itemize}
\newpage

\begin{figure}[h]
\epsfxsize 4.5in
\epsfysize 4.5in
\epsffile{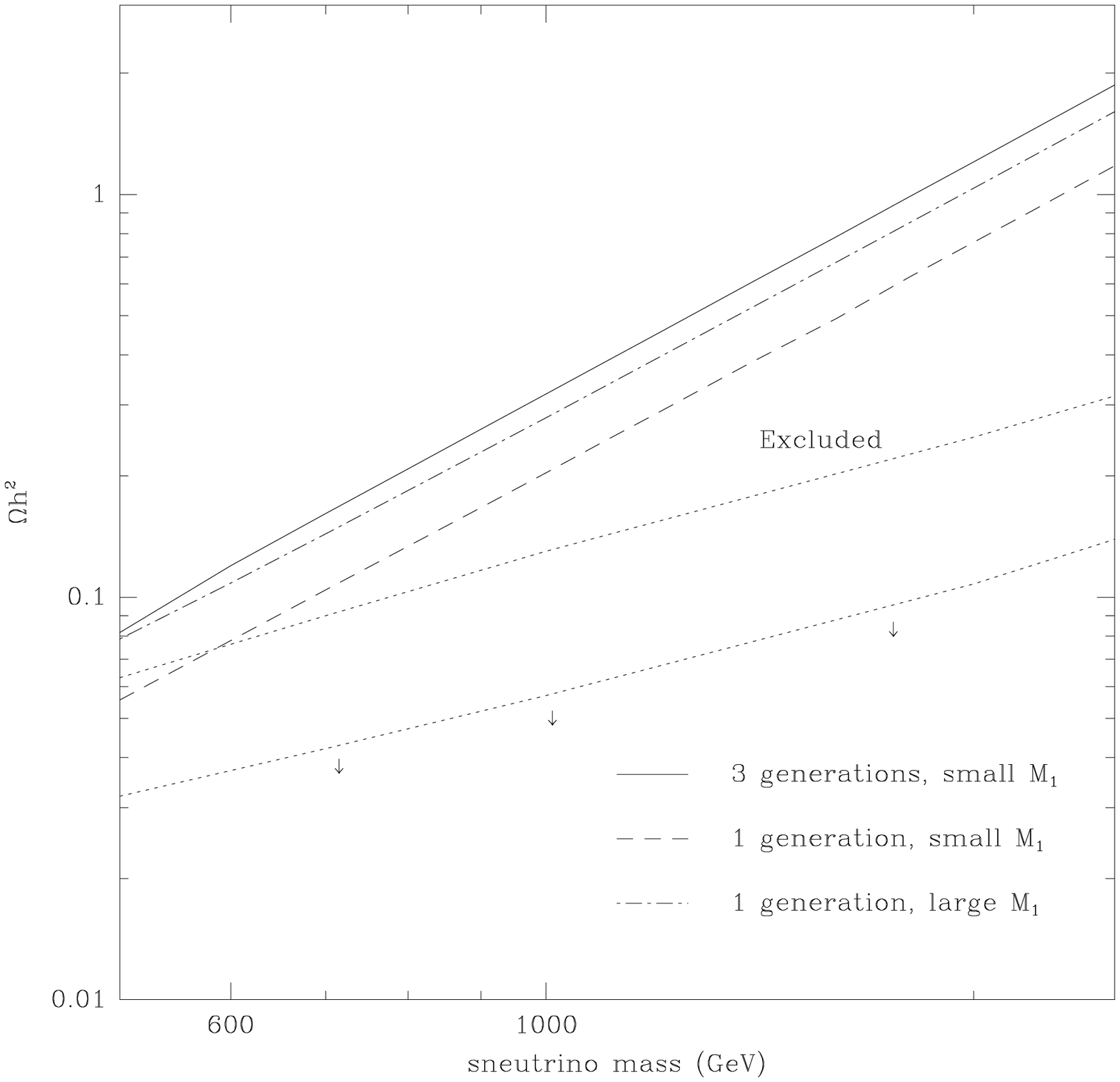}
\end{figure}

\end{document}